\documentclass[aps,prl,preprint,groupedaddress]{revtex4}
\usepackage{graphicx}

\begin{document}

%Title of paper
\title{Neutron resonances in few-body systems and the EOS of neutron star crust}

% repeat the \author .. \affiliation  etc. as needed
% \email, \thanks, \homepage, \altaffiliation all apply to the current
% author. Explanatory text should go in the []'s, actual e-mail
% address or url should go in the {}'s for \email and \homepage.
% Please use the appropriate macro foreach each type of information

% \affiliation command applies to all authors since the last
% \affiliation command. The \affiliation command should follow the
% other information
% \affiliation can be followed by \email, \homepage, \thanks as well.
\author{N. Takibayev
\footnote{E-mail: takibayev@gmail.com}}
%\altaffiliation{Al-Farabi Kazakh National University}
\affiliation{Institute of Experimental and Theoretical Physics, Almaty, Kazakhstan}
\author{K. Kato
\footnote{E-mail: kato@nucl.sci.hokudai.ac.jp}}
%\altaffiliation{Al-Farabi Kazakh National University}
\affiliation{Hokkaido University, Sapporo, Japan}
\author{M. Takibayeva, A. Sarsembayeva and D. Nasirova}
%\altaffiliation{Al-Farabi Kazakh National University}
\affiliation{Al-Farabi Kazakh National University, Almaty, Kazakhstan}
%\noaffiliation

\date{\today}

\begin{abstract}
The effective interactions formed by neutron rescattering between the nuclei fixed in nodes
of the crystalline lattice of neutron star crusts have been considered. In the case of two-body resonances in
neutron-nucleus subsystems new neutron resonances of few-body nature come into existence in the overdense crystal under certain conditions.
The energies and widths of new resonances get additional dependence on the lattice parameters. 
The effective interactions result in nonlinear correction to the equation of state determined by the balance of gravitational, Coulomb and nuclear resonance forces. This leads to resonant oscillations of density in the accordant layers of crusts that are accompanied by oscillations of gamma radiation. 
The phenomena may clarify some processes connected with few-body neutron resonances in neutron star crusts, 
that have influence on the microstructure of pulsar impulses.

\end{abstract}

% insert suggested PACS numbers in braces on next line
\pacs{21.45.+v \and 28.20.-v \and 26.60+c}
% insert suggested keywords - APS authors don't need to do this
\keywords{Neutron resonances \and Few-Body Systems \and Neutron star crusts}

%\maketitle must follow title, authors, abstract, \pacs, and \keywords
\maketitle

% body of paper here - Use proper section commands
% References should be done using the \cite, \ref, and \label commands
\section{Introduction}
% Put \label in argument of \section for cross-referencing
%\section{\label{}}

The specific interaction of neutron with two or more nuclei appears from the quantum problem of scattering of a light particle on the subsystem of heavy particles  ~\cite {bib:Tak1}. 

The neutron-nucleus scattering has a lot of resonances at low energy region extending up to hundreds of $keV$  ~\cite{bib:SM}.
In the few-body system there is a more complicated resonance behavior with a distinctive feature such as the additional  dependence on distances between the nuclei, i.e. on the lattice parameter - $ d $.   

The few-body resonance arises at the certain value of $d = d_r$ ($r=1,2...$), and disappears if the parameter $ d $ becomes somewhat less, or, conversely, more than the resonance value. 
Usually the neutron resonances of elastic scattering are accompanied with resonances in neutron capture channel. And the  elastic and inelastic amplitudes are resonantly enhanced at the same values of the neutron energies and  
$ d_r $ ~\cite{bib:Tak3}.
 
Note that the values of $ d_r $ are much smaller than the size of atoms, but much larger than the size of nuclei. Thus, for neutrons with energies $ E \sim $ 100 keV, the resonances arise when $ d_r \sim $ 50 fm, and for $ E \sim $ 100 eV, $ d_r \sim $ 400 fm. Such distances between the nuclei can not be achieved under normal conditions. They appear only in the overdense crystalline structures of envelopes of neutron stars where mass densities are $ 10^6 \,g\cdot cm^{-3} < \rho < 4 \cdot 10^{11}\,g\cdot cm^{-3}$ for outer crust,  and $ 4 \cdot 10^{11}\,g\cdot cm^{-3} > \rho > 1.4\cdot 10^{14}\,g\cdot cm^{-3}$ for inner crust.  

In the outer crust nuclei are completely stripped, and due to the energy gain of the system form an almost perfect Coulomb crystal in the degenerate electron Fermi gas. The pressure of degenerate electron Fermi gas is opposed to the giant gravitational pressure striving to compress the matter. 
In the lower layers of the outer crust nuclei begin to capture electrons with emission of neutrinos. These reactions lead to the formation of neutron-rich nuclei. Deeper, in the inner crust, neutron-rich nuclei start to emit free neutrons 
 ~\cite{bib:ST}. 

Note that the reactions of electron capture give a lot of new nuclei not only in ground state but also in exited states.   
The excited nuclei interact nonlinearly between each other as the result of overlaps of their wave functions and tunneling effects. Nonleniar interactions stimulate the high harmonic generation, i.e. production of photons with energies many times greater than the excitation energy of a single source ~\cite{bib:SSS,bib:Gan}. 
Therefore, the energetic photons emitted by group of exited nuclei are able to heat the surrounding layers and produce the neutrino-antineutrino pairs interacting with electrons, and even knock the nucleons out from other nuclei. 
It means that free neutrons can appear in the overdense crystal almost immediately after the beginning of electron capture reactions in the corresponding layers of neutron star envelopes.

In this case the resonance forces can change the balance of gravitational and Coulomb forces, which leads to the oscillations in the local layers.

\section{Few-Body Neutron Resonances In Crystalline Structures}
%\label{sec:1}
It is clear that pressures created by the various forces should compensate each other, for example, the balance of gravitational and Coulomb forces gives $P = P_G + P_C = 0$. Denote  $ d_0 $ as the equilibrium value of the lattice parameter in this case. Note that the forces and pressures increase monotonically with the depth in the crystal structure, and the values of $ d_0 $ is monotonically decreasing.

The resonance forces give a new relationship: $P = P_G + P_C + P^{ef}_{res}= 0$. 
Pressure is determined by expression  ~\cite{bib:ST}: $P = \theta^2 \partial (E/\theta)/\partial \theta$,
where $ \theta^{-1} $ is the volume per one baryon, $ E $ is the total energy density of the respective forces.  

The relevant effective energy can be determined with the Hellmann-Feynman relationship: $E^{ef}_n = <\chi_n|V^{ef}|\chi_n>/<\chi_n|\chi_n> $, where $V^{ef}$ is the effective interaction between the lattice nuclei, $ n $ is a neutron. Then, using $ \chi_n $ - the wave functions of a neutron in the potential well the expression can be rewritten as $E^{ef}_n = \xi_n \widetilde{V}^{ef}$, highlighting $ \xi_n $ - the number density of free neutrons in the local layer.  

In general, to determine $ E^{ef} $ as a whole in the structure the sum of the energies of all states $ n $ close to this resonance level must be taken. And besides, it is necessary to make summation over all other resonant levels that appear in this layer.

In turn, $V^{ef}$ can be defined in the problem of neutron scattering on subsystem of few heavy nuclei. Figure 1 shows the resonance behaviour of $V^{ef}$ in the lattice of $^{57}Fe$. And for $n + A_i + A_j  \rightarrow  n + A_i + A_j $ analytical solutions can be obtained in the Born-Oppenheimer approximation with the neutron-nucleus  $ t_i $-matrices taken in the separable or Breit-Wigner form  ~\cite{bib:Tak1}, $ A_i $ is a nuclide, $ i $ - its number in the system.

\begin{figure}
\includegraphics{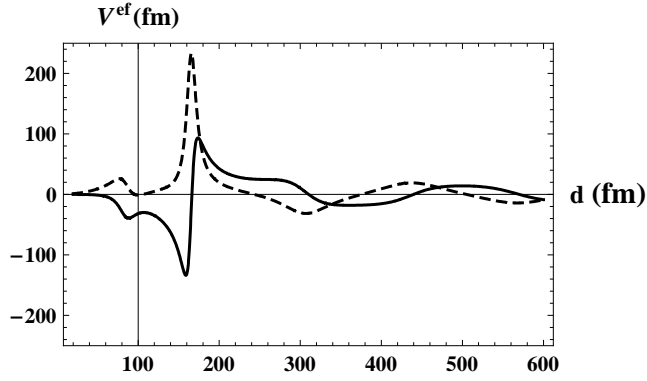}%
% figure caption is below the figure
\caption{$V^{ef}$ for ($n + ^{57}Fe + ^{57}Fe$)-system. The solid and dashed lines correspond to the real and imaginary parts of $V^{ef}$, respectively. Only the two-body ($n+^{57}Fe$)-resonance with energy  $E_R = 52.7 \, keV$ and width $\Gamma = 0.24 \, eV$ is taken into account, the neutron energy $E_0 \approx E_R$.
\label{fig:1}}
%\label{fig:1}       % Give a unique label
\end{figure}

For the pair $ t $-matrix: $ t_i = \bar{\nu}_i ({\bf k}) \eta_i (k_{0}) \nu_i ({\bf k} ') $, relationship with the Breit-Wigner resonances is determined in a simple form $\eta_{i}^{-1} = (E_0 - E_{R_i} + i \Gamma_i /2)$ , где $E_0 = k_0^2/2m$ - the initial neutron energy, $ E_ {R_i}, \Gamma_i $ - the energy and width of neutron-nucleus resonance, and $ \nu_i (k) \simeq \sqrt {\Gamma_i / (4 \pi \, mk)} $ for $ E \simeq E_ {R_i} $. We use the symbolic notation $\nu_i ({\bf k}) = \nu_i (k)\cdot Y_{LM}(\widehat{\bf k})$ and units $ \hbar = 1, c = 1 $, for simplicity. 
Then, the effective interaction between the nuclei is: 
\begin{equation}
V^{ef}_{ij}(k_0;{\bf r, r'}) = C_{ij}\eta_i (k_{0})M_{ij}(k_0;{\bf r, \bf r'}) 
\eta_j (k_{0}) ,\label{eq:1}
\end{equation}
where $ C_ {ij} = \bar {\nu}_i ({\bf k}) \cdot \nu_j ({\bf k}') $, and $ {\bf r} $ and $ {\bf r' }$ are the coordinates of the scattering centers at input and output of a neutron from the system, measured from the point of symmetry between the two nuclei.

The amplitude $M_{ij}(k_0;{\bf r, \bf r'}) = M^+_{ij}(k_0;{\bf r})\delta ({\bf r} + {\bf r'}) + M^-_{ij}(k_0;{\bf r})\delta ({\bf r} - {\bf r'})$ and its components
\begin{equation}
M^+_{ij}(k_0;{\bf r}) = \frac{1}{D_{ii}}J_{ij}(k_0;{\bf r}) \ , \ \ \
M^-_{ii}(k_0;{\bf r}) = \frac{1}{D_{ii}}J_{ik}(k_0;{\bf r})\eta_k (k_{0})J_{ki}(k_0;{-\bf r}).  
\end{equation}
The elements of the matrix $D_{ij} = \delta_{ij} - J_{ik}(k_0;{\bf r})\eta_k(k_{0})J_{kj}(k_0;{-\bf r})\eta_j (k_{0})$, where by definition  $J_{ii} = 0$, and  
$J_{ij}(k_0;{\bf r})$ is the Fourier transform of the Born interaction in the three-body system:
\begin{equation}
J_{ij}(k_0;{\bf r}) = 2m \int d{\bf k} \exp(i{\bf kr}) \frac{\nu_i ({\bf k})\bar{\nu}_j ({\bf k})}{k_0^2 - k^2 + i0} \  .
\end{equation}
The zeros of the determinant $ D = 0 $ in the complex plane $ k_0 $ correspond to resonant states of the system and $d_r = 2|{\bf r}_r|$.

Figure 2 shows the deviation of the equilibrium values of $ d_0 $ - in the absence and $ d_{0, res} $ in the presence of neutron resonances in the local layers of the crystalline structure. The resulting local oscillations give reason to assume that the radiation accompanying the structural neutron resonances will also have oscillatory behavior.

The value of $ P^{ef}_{res} $ is almost everywhere close to zero, and becomes significant near the $d \approx d_{res}$. That is, each nuclide and each resonant level in the ($ n, A $)-subsystem correspond to a certain segment of layers, where $ d \approx d_{res} $ and, consequently, the density $ \rho \approx \rho_{res} $. This is an important characteristic of the resonance strength. 

The combined effect of forces gives a new equilibrium value of $ d_{0, res} $, which may not have a monotonic behavior.

It is interesting that $ d_{0, res} $ may appear in the region where the resonance interaction becomes vanishingly small. In this case, the balance will settle without the participation of resonance forces, and the density of local layers will tend to return to its original value: $d \rightarrow d_0$. 
However, as before, if $ d_0 $ comes close to $ d_{res} $, the resonance forces will rise again and change the balance of power, again leading to $ d_{0, res}$. 

This mechanism can result in local oscillations of density and generate a radiation in this area with form of the comb, which may explain the complicated microstructure of the impulses of some pulsars.

\begin{figure}
\includegraphics{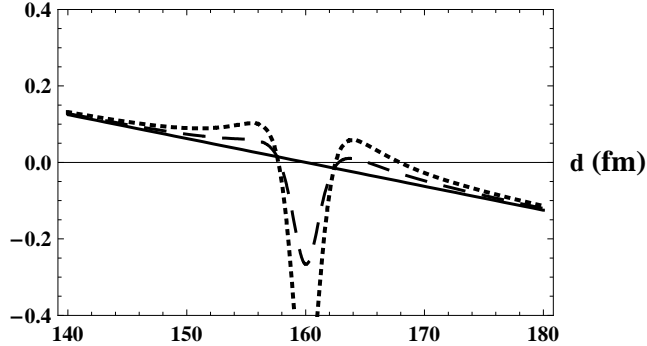}%
% figure caption is below the figure
\caption{The sum of pressures in relative terms: $(P_G + P_C + P^{ef}_{res})/P_C$. Solid line corresponds neutron density $ \xi_n = 0$, i.e. $P^{ef}_{res}= 0$, dashed line - $ \xi_n = 6\cdot 10^{-5} fm^{-3}$ and the dotted line - $ \xi_n = 1.44 \cdot 10^{-4} fm^{-3}$.
\label{fig:1}}       % Give a unique label
\end{figure}

\begin{acknowledgements}
This work was supported by the MES of Republic of Kazakhstan (the Grant No. 1133/SF). The authors thank the participants of theoretical physics seminar of Hokkaido University and RCNP of Osaka University. N.Takibayev thanks the Hokkaido University and RCNP of Osaka University for their hospitality and support. 
%If you'd like to thank anyone, place your comments here
%and remove the percent signs.
\end{acknowledgements}

\end{document}